%% file: sample-authordraft.tex
\documentclass[manuscript]{acmart}

\AtBeginDocument{%
  \providecommand\BibTeX{{%
    \normalfont B\kern-0.5em{\scshape i\kern-0.25em b}\kern-0.8em\TeX}}}

\copyrightyear{2022}
\acmYear{2022}
\setcopyright{acmcopyright}\acmConference[FAccT '22]{2022 ACM Conference on Fairness, Accountability, and Transparency}{June 21--24, 2022}{Seoul, Republic of Korea}
\acmBooktitle{2022 ACM Conference on Fairness, Accountability, and Transparency (FAccT '22), June 21--24, 2022, Seoul, Republic of Korea}
\acmPrice{15.00}
\acmDOI{10.1145/3531146.3534628}
\acmISBN{978-1-4503-9352-2/22/06}

\usepackage{csquotes}

\begin{document}

\title{The Conflict Between Explainable and Accountable Decision-Making Algorithms}

\author{Gabriel Lima}
\email{gabriel.lima@kaist.ac.kr}
\orcid{0000-0002-2361-350X}
\affiliation{%
  \institution{School of Computing, KAIST \& Data Science Group, IBS}
  \country{Republic of Korea}
}

\author{Nina Grgi\'{c}-Hla\v{c}a}
\email{nghlaca@mpi-sws.org}
\orcid{0000-0003-3397-2984}
\affiliation{%
  \institution{Max Planck Institute for Software Systems \& Max Planck Institute for Research on Collective Goods}
  \country{Germany}
}

\author{Jin Keun Jeong}
\email{jkjeong@kangwon.ac.kr}
\affiliation{%
  \institution{Kangwon National University}
  \country{Republic of Korea}
}

\author{Meeyoung Cha}
\email{mcha@ibs.re.kr}
\orcid{0000-0003-4085-9648}
\affiliation{%
  \institution{Data Science Group, IBS \& School of Computing, KAIST}
  \country{Republic of Korea}
}

\renewcommand{\shortauthors}{Lima \emph{et al.}}

\begin{abstract}
    \input{content/0abstract}
\end{abstract}



\maketitle

\section{Introduction}

\input{content/1intro}

\section{Background}

\input{content/2background}

\section{The Conflict Between Explainability and Responsibility}

\input{content/3tradeoff}

\section{Implications}

\input{content/4implications}

\section{Concluding Remarks}

\input{content/5conclusion}

\begin{acks}
This work was supported by the Institute for Basic Science (IBS-R029-C2) and the National Research Foundation of Korea (NRF-2017R1E1A1A01076400). 
\end{acks}

\bibliographystyle{ACM-Reference-Format}
\bibliography{sample-base}

\end{document}

%% file: content/0abstract.tex
Decision-making algorithms are being used in important decisions, such as who should be enrolled in health care programs and be hired. Even though these systems are currently deployed in high-stakes scenarios, many of them cannot explain their decisions. This limitation has prompted the Explainable Artificial Intelligence (XAI) initiative, which aims to make algorithms explainable to comply with legal requirements, promote trust, and maintain accountability. This paper questions whether and to what extent explainability can help solve the responsibility issues posed by autonomous AI systems. We suggest that XAI systems that provide post-hoc explanations could be seen as blameworthy agents, obscuring the responsibility of developers in the decision-making process. Furthermore, we argue that XAI could result in incorrect attributions of responsibility to vulnerable stakeholders, such as those who are subjected to algorithmic decisions (i.e., patients), due to a misguided perception that they have control over explainable algorithms. This conflict between explainability and accountability can be exacerbated if designers choose to use algorithms and patients as moral and legal scapegoats. We conclude with a set of recommendations for how to approach this tension in the socio-technical process of algorithmic decision-making and a defense of hard regulation to prevent designers from escaping responsibility.

%% file: content/1intro.tex
Artificial Intelligence (AI) is now used in a wide range of situations, from low-stakes scenarios like entertainment~\cite{silver2017mastering} to high-stakes life-or-death decisions like selecting who should be prioritized for medical help~\cite{obermeyer2019dissecting}. Extensive research has inquired whether algorithmic decision-making has negative implications for society. Studies have observed, for instance, algorithmic bail decisions to be racially biased~\cite{propublica_story}, discussed how AI systems\footnote{We use AI systems and decision-making algorithms interchangeably. This is in line with the Explainable AI (XAI) literature. We note that most decision-making algorithms would be labeled as AI systems regardless of their complexity or autonomy.} used for hiring decisions could embed biases~\cite{barocas2016big}, and found online advertisement to discriminate against women~\cite{datta2015automated}.

A major problem with most decision-making algorithms is their opacity. Most algorithms are black boxes that do not offer explanations for their decisions, recommendations, or processing~\cite{pasquale2015black}. This limitation has been a central motivation for developing Explainable Artificial Intelligence (XAI), which proposes to make algorithms explainable by ``making [their] functioning clear and easy to understand''~\cite{arrieta2020explainable}. In the context of algorithmic decision-making, XAI creates models whose behavior can be easily understood (i.e., those that are transparent) or that can explain their behavior after a decision (e.g., by providing post-hoc explanations). Calls for XAI have become widespread in industry, academia, and policymaking~\cite{jobin2019global}.

The XAI field aims to create systems that facilitate attributing responsibility to human agents involved in their development and deployment. Assigning responsibility for algorithmic decisions has been widely debated to be a difficult task due to the existence of a responsibility gap~\cite{matthias2004responsibility,asaro2016liability}. As argued by~\citet{robbins2019misdirected}, explainable systems would maintain meaningful human control, allowing responsibility to be traced back to designers,\footnote{We refer to developers and designers as the \emph{collective} agents that encompass programmers, executives, and any other entity involved in the design and development of decision-making algorithms. More specifically, we turn our attention to corporations that may develop decision-making algorithms.} users, and patients (i.e., those subjected to algorithmic decision-making). That is not to say that XAI is only put forward to deal with responsibility issues. Explanations can, for instance, also be used to comply with legal requirements, promote trust in decision-making algorithms, and assess their accuracy~\cite{langer2021we}.

We argue in this paper that XAI is not a panacea to the plethora of responsibility issues that autonomous decision-making algorithms entail. We focus our discussion on AI systems that are designed to make consequential decisions and can provide explanations afterwards, i.e., algorithms that are explainable in a post-hoc manner. While we agree that explainable systems are necessary for the responsible deployment of algorithmic decision-making, we show how XAI's post-hoc explanations may be at odds with the public's understanding of AI systems' agency and blameworthiness. Furthermore, we discuss how those who are subjected to algorithmic decisions (i.e., patients) may be perceived as having meaningful human control over XAI systems and illustrate how this impression is false and does not translate to true empowerment over algorithms.

Considering blame as a response to the reasons upon which an agent has acted~\cite{scanlon2009moral}, post-hoc explainable algorithms may be perceived as actors that can explain the reasons behind their decisions and thus as blameworthy. Explainable AI systems may also be viewed as more capable and intentional than their opaque counterparts, resulting in higher levels of blame~\cite{malle2014theory,malle2006intentionality}. This impression obscures the responsibility of human agents in algorithmic decision-making and shifts laypeople's moral judgments towards machines, potentially influencing policymakers and hindering the adoption of beneficial AI technologies~\cite{cave2018portrayals,bonnefon2020moral}.

Motivated by the concern that developers could launder their agency for the deployment of autonomous systems~\cite{rubel2019agency} and implement superficial ethical measures to avoid regulation~\cite{floridi2019translating}, we show how they could use XAI to create a false sense of understanding and control for patients. We illustrate this misleading impression with research showing that algorithmic explanations are often nonsensical and leave individuals with no real control~\cite{rudin2019stop}. XAI systems can also be used to deceive patients, even those trained in AI-related areas~\cite{ehsan2021explainability}, creating moral and legal scapegoats.

By illustrating how the responsibility for explainable systems might be blurred, we contribute to the literature by analyzing the responsibility gap posed by autonomous systems with a novel and critical perspective on the XAI field. We conclude with a call for interpretable systems, which would emphasize developers' responsibilities throughout the development and deployment of decision-making algorithms. Finally, we discuss how current regulatory approaches fail to address the conflict between explainability and responsibility and offer potential solutions.

%% file: content/2background.tex
\subsection{Explainable AI}

Algorithms are used to assist human judges in bail decisions~\cite{propublica_story}, decide which patients should be prioritized for medical assistance~\cite{obermeyer2019dissecting}, evaluate job applicants~\cite{barocas2016big}, and in many other applications. Algorithmic decision-making has become widespread in society, and much research has been devoted to understanding its benefits and drawbacks. A common criticism of most models used for making these life-changing decisions is that they are inscrutable black boxes~\cite{selbst2018intuitive}. Users, patients, and even designers do not understand how algorithms make decisions, making it impossible to backtrack their decision-making process.

The widespread use of consequential decision-making algorithms has made understanding how they work necessary. Explainable Artificial Intelligence (XAI), a field committed to increasing people's understanding of decision-making algorithms, arose from this need. As defined by Arrieta et al.~\cite{arrieta2020explainable}, an explainable system is ``one that produces details or reasons to make its functioning clear or easy to understand'' for a specific audience, be it users, designers, patients, or policymakers.

Efforts to develop explainable systems are often categorized into two groups: they either propose \emph{transparent} models or build techniques to assist black box models to explain their behavior after a decision (referred to as \emph{post-hoc explainability)}~\cite{lipton2018mythos}. Algorithms are transparent when a human can simulate its functioning (i.e., the model is simulatable), explain each part of the model (decomposable), and follow its decision-making process. Post-hoc explainability, on the other hand, is supported by strategies and models that explain decisions for any given input.

A linear regression is a typical example of a transparent model.
Variables can be human-readable, and people can simulate the model if it is not unnecessarily complex. There are countless examples of post-hoc explainability, including techniques for simplifying an algorithm's specific decision (e.g., LIME~\cite{ribeiro2016should}), creating human-understandable visualizations (e.g.,~\cite{selvaraju2017grad}), and presenting counterfactual examples (e.g.,~\cite{wachter2017counterfactual}). Post-hoc explanations present extra information, such as which feature of an input had the greatest impact on the final decision or similar examples that might have resulted in a different determination (i.e., a counterfactual). This paper focuses on post-hoc explanations and how they may conflict with accountability.

\subsection{Widespread Calls for Explainable AI}

In a study of the global guidelines referring to the ethics of AI, ~\citet{jobin2019global} found explainability (and similar notions like transparency) to be the most prominent principle across the efforts to promote the responsible development and deployment of AI. Although not restricted to algorithmic decision-making, several guidelines proposed explainability for this specific context~\cite{council2017statement,floridi2018ai4people}. Industry leaders have also called for research in the field of XAI. For instance, Microsoft's CEO Satya Nadella defended the use of transparent systems, ``AI must be transparent. [...] People should have an understanding of how the technology sees and analyzes the world''~\cite{nadella}. Researchers have also observed an exponential increase in the number of articles addressing XAI being published in academic venues in the last few years~\cite{arrieta2020explainable}, demonstrating the role of academia in the development of explainable algorithms.

Scholars have argued for explainable systems based on a variety of premises. Felzmann et al.~\cite{felzmann2019transparency} justified the research agenda to promote the acceptance of AI systems. Mittelstadt et al.~\cite{mittelstadt2019explaining} defended explainability as a form of promoting trust and verification in algorithmic decision-making. Burrell~\cite{burrell2016machine} argued in favor of explainable systems as a form of assessing fairness. Systems that are understandable for stakeholders could also contribute to their privacy~\cite{arrieta2020explainable}.

The development of XAI has also been motivated by legal requirements~\cite{bibal2020legal}. Existing laws, such as those granting consumers access to their credit score, require algorithms to explain their decisions~\cite{selbst2018intuitive}. The ``right to explanation'' included in the European General Data Protection Regulation (GDPR) is a good example of how politicians could enforce explainability~\cite{goodman2017european}. While interpretations of the law are yet to take place in courts, GDPR seems to entitle patients the right to ask for explanations concerning the logic used by algorithms that make consequential decisions~\cite{selbst2018intuitive}.

Although the introduction to XAI presented above portrays it as a panacea to the opacity posed by algorithmic decision-making, the field has also been subjected to serious criticism. Scholars have raised concerns about how explainable systems may give designers unwarranted control over the information that patients and users receive~\cite{barocas2020hidden}. \citet{kasirzadeh2021use} criticized counterfactuals as an appropriate approach to XAI as it might require incoherent social theories, e.g., by disregarding that some social categories are immutable. \citet{de2018algorithmic} argued that full transparency should be avoided due to potential negative consequences, such as loss of privacy. Robbins~\cite{robbins2019misdirected} showed that requiring explainability of all AI systems is misguided and suggested that a principle of explainability should be connected to specific decisions rather than the technology.

Building upon this previous work, our novel critique of XAI relies on its conflict with one issue it seeks to solve: the responsibility gap posed by autonomous systems. Scholars defend that XAI contributes to the report of an algorithm's negative impacts and auditability~\cite{arrieta2020explainable}. Explainability has often been framed as a necessary condition for accountable AI systems (e.g.,~\cite{miller2019explanation}). Similarly, Robbins~\cite{robbins2019misdirected} argued that explainable systems are primarily for maintaining meaningful human control and responsibility. Before delving deeper into how XAI could clash with accountability, we introduce the notion of the responsibility gap and how scholars have addressed it below.

\subsection{Responsibility Gaps}

Algorithms decide who is automatically enrolled in health care programs~\cite{obermeyer2019dissecting} and determine which job opportunities are shown online to job seekers~\cite{datta2015automated}. If these systems are found to discriminate against a certain social group, who should be held responsible for any harm caused by this decision? One may claim that the designers should have foreseen this risk during the development phase, thereby making them responsible. Another possibility is to hold those who use these systems accountable for their decision to use algorithms in the real world. Because these algorithms examine past data to make decisions, persons who are subjected to algorithmic decisions (i.e., patients) may be viewed as important actors who influenced the decision. The dilemma of which of these entities is an appropriate subject of responsibility arises for autonomous and adaptive algorithms (e.g., AI systems), creating a responsibility gap.

The debate surrounding the responsibility gap was initiated by ~\citet{matthias2004responsibility}, who argued that autonomous and self-learning machines threaten the necessary conditions for holding someone responsible. The degree of control that operators, users, and manufacturers have over a machine's behavior is challenged by high levels of autonomy. Any attempts to predict the behavior of systems built to constantly learn and adapt to new contexts are limited, conflicting with epistemic requirements for accountability. This lack of control and knowledge creates a responsibility gap, under which no one is a suitable subject of responsibility. This gap refers to various aspects of responsibility, ranging from who is to blame (e.g., blameworthiness) to who should be held accountable for algorithmic harm to the public (e.g., public accountability)~\cite{de2021four}.

\subsection{Bridging the Responsibility Gap}


How may society bridge the responsibility gap? Some scholars argue that algorithms, no matter how intricate, autonomous, or self-learning, are just human tools~\cite{bryson2010robots}. They contend that an algorithm's behavior should be understood as a collection of its users' and designers' decisions and intentionality rather than as intentional actions per se~\cite{johnson2006computer}. These viewpoints maintain human agents as the subjects of responsibility for legal coherence~\cite{bryson2017and}.

\citet{coeckelbergh2020artificial} takes a different approach and proposes to ground responsibility on the patients of algorithms' actions to circumvent the responsibility gap. This approach proposes that those who use and develop AI systems should take on forward-looking responsibilities by focusing on how algorithms could harm those subjected to it. \citet{champagne2015bridging} present a similar proposal in the context of automated warfare, arguing that a ``person of sufficiently high standing could accept responsibility for the actions of autonomous robotic devices.'' This perspective appears to be the approach taken by some self-driving car manufacturers, who have pledged to bear responsibility for any accidents caused by their machines~\cite{volvo}.

Another approach for bridging the responsibility gap is to hold AI systems accountable. \citet{stahl2006responsible}, for instance, claims that machines could be held (quasi-)responsible, fulfilling several social goals derived from moral responsibility. \citet{coeckelbergh2009virtual} discusses how AI systems could be held ``virtually'' responsible to the extent that they appear to be morally responsible. These views are related to the proposal to extend legal personhood and responsibility to AI systems~\cite{turner2018robot,gordon2020artificial}; yet, such a proposal has encountered much opposition~\cite{bryson2017and,solaiman2017legal}.

The proposal to hold algorithmic systems responsible has been controversial. Punishing an AI system, for instance, would be meaningless because it is incapable of suffering or carrying culpability~\cite{sparrow2007killer,danaher2016robots}. Others have argued that holding AI systems accountable requires metaphysical properties that existing actors lack, such as consciousness~\cite{himma2009artificial} and sentience~\cite{torrance2008ethics}.

A less controversial view proposes to view AI systems as human-AI collaborations. For instance, Nyholm~\cite{nyholm2018attributing} argues that machines' agency should be viewed as a collaborative effort between them and human actors, in which the latter ``initiate, supervise, and manage the agency'' of the former. Such a viewpoint argues that humans should remain the locus of responsibility. Another comparable view defends a form of joint responsibility, in which humans and autonomous algorithmic systems share responsibility~\cite{gunkel2017mind,hanson2009beyond}.

%% file: content/3tradeoff.tex
\begin{figure*}[t!]
    \centering
    \includegraphics[width=0.9\linewidth]{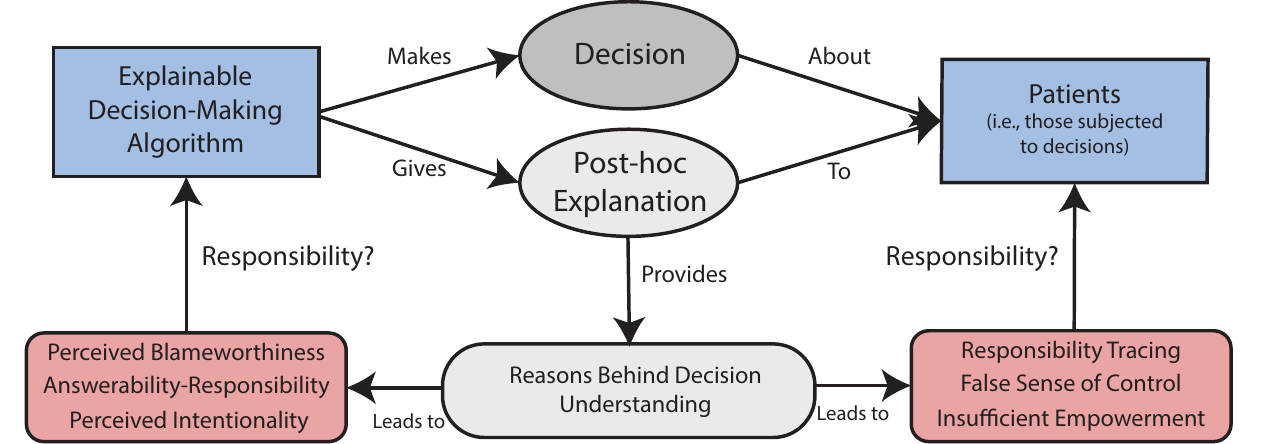}
    \caption{Our main points are summarized in this diagram. 
    Decision-making algorithms that provide post-hoc explanations can be viewed as blameworthy agents that should be held accountable for their decisions; we explain how this perception can be explained by increased attributions of intentionality and capacity. 
    Explainable systems can also result in incorrect attributions of responsibility to those subjected to algorithmic decisions (i.e., patients). Explainable systems give the impression of confidence and empowerment, implying that patients should bear some responsibility. This notion is incorrect and could be exploited by designers attempting to escape responsibility for algorithmic decision-making.}
    \label{fig:img}
\end{figure*}

In this paper, we focus on XAI systems that make consequential decisions, such as those used in the medical~\cite{obermeyer2019dissecting}, hiring~\cite{nytresume}, and financial~\cite{themarkuphousing} domains. Despite the fact that most AI systems are designed to work alongside human decision-makers, algorithms may nonetheless operate as decision-makers in practice. For instance, AI determines which job applications should be later evaluated by human employers, having a direct impact on employment prospects~\cite{nytresume}. These AI systems are often marketed as decision-making tools, influencing how they are used in the real world~\cite{themarkuphousing}. The responsibility gap becomes significant in these scenarios because human decision-making is undermined or even dismissed, blurring human responsibility.

We argue that XAI poses a unique challenge and may contribute to the responsibility gap (see Figure~\ref{fig:img} for a summary of the main arguments). First, we show how AI systems that provide post-hoc explanations can be viewed as blameworthy agents, obscuring human responsibility. Second, we argue that explainable algorithms create the false impression that patients have meaningful human control, leading to incorrect attributions of responsibility. Third, we look at how designers' control over XAI systems allows them to use algorithms and patients as scapegoats, escaping responsibility.

\subsection{Explainable Systems as Responsible Agents}
\label{sec:respAI}

What does it mean to blame someone? Scanlon~\cite{scanlon2009moral} has proposed a distinct interpretation of blame, in that judgments of blameworthiness are based on one's assessment of others' reasons behind their intentions and attitudes that go against the standards of their relationship, i.e., their ``social contract.'' To blame someone is to respond to this impairment by modifying one's views on their relationship with the blamee. 

Based on Scanlon's interpretation, we suggest that decision-making algorithms may be deemed blameworthy if they appear to provide the reasons for their decisions. When someone requests an explanation for an algorithmic decision, they expect to hear the reasons for that determination~\cite{doshi2017accountability}. As a result, AI systems that can provide post-hoc explanations may be deemed blameworthy if their reasons go against what is expected from them. It is worth noting that one of the issues covered by the responsibility gap is blameworthiness~\cite{de2021four}.

There is evidence that people blame AI systems and algorithms when they cause harm, regardless of their explainability. For instance, people blame algorithms when they make life-or-death decisions about who should live or die in a moral dilemma~\cite{malle2019ai}. Another study found that AI systems described as autonomous are blamed to a similar extent to human agents~\cite{furlough2019attributing}. Research has also shown similar results in several scenarios, ranging from medicine~\cite{lima2021punish} to autonomous vehicles~\cite{franklin2021blaming}. In the context of algorithmic decision-making, AI systems making bail decisions are blamed similarly to human judges~\cite{lima2021human}. In this paper, we do not argue that explainability alone affects blameworthiness. Instead, we argue that explainable systems may be subjected to higher levels of blame, conflicting with its aim to trace responsibility back to human actors.

One possible criticism of the argument above is that blameworthiness does not encompass all notions of moral responsibility. Moral philosophers have highlighted the pluralistic component of moral responsibility (e.g., \cite{van2015moral}). For instance, Shoemaker~\cite{shoemaker2011attributability} differentiates between three different notions of moral responsibility: attributability, answerability, and accountability. While attributability does not seem to apply to AI systems because they rely on evaluations of an agent's character, the latter two notions pose the same contradiction we defended above. First, Shoemaker defines responsibility-as-accountability similarly to Scanlon's blameworthiness, which we discussed above. Second, Shoemaker poses that to be answerable relies on an agent's ability to cite the reasons upon which their actions are grounded. Hence, AI systems that can explain the reasons behind their decisions could also be perceived as morally responsible with respect to answerability. We do not claim that XAI systems will be perceived as responsible according to all definitions of moral responsibility; nevertheless, this perception could obscure the moral responsibility of human agents in algorithmic decision-making.

Empirical evidence also supports the role of reasons in assigning blame and moral responsibility. Blame can be mitigated if the agent has justifiable reasons; in contrast, blame can be increased if the agent bases their actions on improper or immoral reasons~\cite{malle2014theory}. The perception of an agent acting upon reasons could lead to higher perceived intentionality. Beliefs and desires, which can be given as reasons for an action, are crucial components of the folk conception of intentionality~\cite{malle1997folk}. Explainable algorithms' decisions could be perceived as intentional, as suggested by prior research showing that humans adopt an intentional stance towards non-human entities~\cite{perez2020adopting} and robots~\cite{marchesi2019we}. Most crucially to our argument, prior work has shown the pivotal role of intentionality in judgments of blameworthiness~\cite{malle2006intentionality}; if XAI systems are perceived as intentional, they are likely to be deemed blameworthy.

Explainable AI systems could also be perceived as blameworthy without attributions of intentionality. Malle et al.'s theory of blame~\cite{malle2014theory} proposes that agents who are perceived to have the capacity to prevent or contribute to an event are also attributed substantial levels of blame. People view XAI systems are largely capable, as evidenced by research showing that individuals overrely on algorithmic recommendations~\cite{bansal2020does,jacobs2021machine}, partly because they overtrust the explanations provided by these systems~\cite{bansal2019beyond}. Explainable systems give the impression that they can make good decisions, and people blindly follow their decisions. This sense of capacity can also result in heightened blame attributions.

All of the research presented thus far suggests that decision-making algorithms could be seen as morally responsible if they can explain their decisions. Therefore, people may hold explainable algorithms responsible for their decisions, even though they may not be appropriate subjects of responsibility~\cite{sparrow2007killer,danaher2016robots,himma2009artificial,torrance2008ethics}. XAI aims to trace responsibility back to human agents, but it may instead shift responsibility to the algorithms it aims to explain.

\subsection{Patients as Responsible Agents}
\label{sec:respUser}

The concept of meaningful human control was proposed as a crucial precondition for maintaining humans ``in control, and thus morally responsible'' for algorithmic decision-making~\cite{santoni2018meaningful}. \citet{santoni2018meaningful} identified two necessary conditions for algorithms to remain under meaningful human control: \emph{tracking} and \emph{tracing}. The tracking condition demands decision-making algorithms to track a human agent's moral reasons relevant to the decision. The tracing condition, on the other hand, requires that someone comprehend the algorithm's capabilities as well as any of their real-world consequences. We observe a conflict between the latter condition of tracing and XAI systems.

If algorithms are explainable in a post-hoc manner, it is expected that patients understand their decisions. Explanations are designed to make algorithms explainable to a specific audience~\cite{arrieta2020explainable}. By design, post-hoc explainable AI aims to make an algorithm more understandable to patients, i.e., those interacting with them in a post-hoc manner. These explanations may thus shift perceived control from designers to those who are subjected to algorithmic decision-making.

Some of the methods presented by XAI researchers aim to empower the patients of algorithmic decision-making. For instance, Wachter et al.~\cite{wachter2017counterfactual} proposed counterfactual explanations as a form of putting the recipients of decisions made by algorithms ``under control.'' They argued that explanations---and more specifically counterfactual explanations---help patients understand why particular decisions were made, empower them to contest decisions with which they disagree, and provide courses of action they can take to achieve a different result in the future. It is worth noting that all of these goals are geared towards giving patients a better grasp of and control over an algorithm's decision-making capabilities. These approaches seem to portray patients as satisfying the tracing condition of meaningful human control. However, is this sense of control meaningful?

\citet{rudin2019stop} criticizes post-hoc explanations, claiming that current approaches could create a false sense of understanding among those subjected to algorithmic decision-making. Rudin explains how saliency maps, which highlight the parts of an image that played a major role in a classification task, do not adequately describe how the model processes those parts identified as salient. Instead, this type of explanation just provides a false sense of understanding. Another example of how explanations could create a misleading sense of control is by giving explanations that are nonactionable~\cite{kasirzadeh2021use}. Algorithms that tell patients they were not granted a loan because they do not have enough collateral, for instance, do not provide them with useful information.\footnote{Some research covers how actionable explanations could be created~\cite{joshi2019towards,venkatasubramanian2020philosophical,ustun2019actionable}; however, this field of inquiry still faces many challenges~\cite{verma2020counterfactual}, such as how to evaluate explanations with patients or account for hidden features.} Although post-hoc explanations appear to empower patients by giving them control over decision-making algorithms, this empowerment is not meaningful.

Empirical research has also shown that patients misuse AI explanations, even when they have an AI-related background~\cite{kaur2020interpreting,ehsan2021explainable}. Algorithmic explanations can potentially deceive individuals who receive them, both intentionally and unintentionally~\cite{ehsan2021explainability}, tricking them into doing things without having their interests in mind. In conclusion, this sense of control and understanding is insufficient for grounding patients as responsible for algorithmic decision-making.

We have suggested that algorithms and patients will be perceived as responsible for algorithmic decisions, but it is unclear whether this view has real-world implications. Legal systems could choose to disregard the public opinion by holding developers accountable regardless of whom laypeople choose to blame or consider morally responsible for algorithmic harm, thus bridging the responsibility gap. However, doing so might hurt the adoption of possibly beneficial AI systems~\cite{bonnefon2020moral}. Detaching legal responsibility from its folk conception might create a ``law in the books,'' which is unfamiliar to the people whose behavior legal systems aim to regulate~\cite{brozek2019can}. According to empirical studies, people's reactions to algorithmic harm conflict with current legal systems~\cite{lima2021punish} and adopting liability models that clash with laypeople's views may hurt the adoption of autonomous vehicles~\cite{liu2021psychological}.

\citet{saetra2021confounding} has suggested that the responsibility issues posed by AI systems have inherent tradeoffs that should be assessed openly through political deliberation. Scholars have noted how laypeople's risk perceptions impacted past regulation of emerging technologies, such as genetically modified organisms (GMOs), and it is expected that laypeople will influence how AI is regulated~\cite{cave2018portrayals}. Policymakers should be aware of the possible backlash caused by policies that go against public expectations~\cite{awad2020crowdsourcing}. In conclusion, people's perceptions of who should be held responsible for algorithmic harm have real-world implications for the development, deployment, and regulation of decision-making algorithms.

\subsection{Designers as Responsible Agents}

All the conflicts discussed above do not seem to pose a problem if designers take responsibility for the decisions made by AI systems. Designers could take responsibility for whatever harm these systems may cause, regardless of whether they are explainable. Such an approach was proposed by Champagne and Tonkens in the context of automated warfare~\cite{champagne2015bridging}. Even if human actors do not meet the necessary conditions for backward-looking attributions of responsibility, they can take proactive responsibility for future consequences.

\citet{van2021responsible} discussed a similar approach in the context of ``responsible robotics.'' While not directly related to algorithmic decision-making, their research emphasizes the importance of a framework that carefully considers the roles of all stakeholders throughout the development and deployment of autonomous agents. More crucially, their approach emphasizes the role of human agents in designing and deploying these systems, rather than considering algorithms (or robots) responsible. If algorithms are held responsible, they may serve only as liability shields~\cite{bryson2017and}.

In a similar vein, \citet{johnson2006computer} argued that responsibility gaps will not arise from the technological aspects of AI systems but can only exist as a result of deliberate decisions. Such gaps will only emerge if humans developing and deploying AI systems fail to create systems that circumvent the responsibility gap, making the development of AI systems that designers can control and understand imperative.

\subsection{Algorithms and Patients as Scapegoats}

We agree that designers should take responsibility for the decisions made by their algorithms and be held to account if something goes wrong. First, designers often have the most assets, placing them in the best position to compensate anyone who has been harmed by these systems and thus accomplishing the primary purpose of holding entities responsible under civil law~\cite{prosser1941handbook}. Considering that designers will profit from the deployment of AI systems, they could ensure that those harmed are compensated~\cite{vcerka2015liability}. Second, if designers are not held responsible, they may continue developing unsafe systems in pursuit of increased profits. Being able to hold someone responsible is critical for promoting cooperation by deterring selfish behavior~\cite{fehr2002altruistic}. However, XAI systems could prove to be a tool for designers who willingly choose to shift perceived responsibility away from themselves. As previously stated, laypeople's perceptions of who is responsible have real ramifications for policymaking and the adoption of XAI systems.

Designers are unlikely to take responsibility for algorithmic decision-making to the extent that is necessary. \citet{rubel2019agency} highlighted how ``using algorithms to make decisions can allow [...] persons to distance themselves from morally suspect actions'' by shifting the responsibility for a decision to the algorithm through agency laundering, as the authors call it. The development of explainable decision-making algorithms creates a series of tools for designers to obscure their involvement and shift responsibility to the system itself. Given that explainable algorithms may be viewed as morally responsible, designers could intentionally emphasize the algorithm's role in the decision-making process. XAI systems could thus become apparent rational and intentional responsibility scapegoats. It is worth noting that explainable decision-making algorithms can be seen as morally responsible regardless of their designers' willingness to participate in agency laundering. Nonetheless, this conflict between accountability and explainability can be aggravated if designers choose to take advantage of this misleading perception.

We have argued that some types of explanations, e.g., counterfactuals, could shift perceived control of a decision-making algorithm from its designers to its patients. Following the same argument presented above concerning agency laundering, designers could highlight the role of patients in the decision-making process to obscure their own responsibility. Design methods that apparently empower patients through explanations can shift perceived responsibility for harmful outcomes to those receiving the decisions and explanations.

Explainability gives designers an exceptional degree of power by allowing them to choose what kind of information is delivered as explanations to patients~\cite{barocas2016big}. As a result, designers can prioritize their interests over the well-being of individuals on the other side of the decision-making process. The concept of "dark patterns" in user interface design~\cite{brignull2015dark} exemplifies this power relation, in which designers can intentionally deceive users without their best interests in mind. These dark patterns can also be extended to XAI~\cite{ehsan2021explainability}, shifting responsibility to patients.

Explainability could also be used as a form of ethics washing. Explainable systems could be implemented as a superficial ethical measure to avoid necessary regulation~\cite{floridi2019translating}. Given the numerous legal requirements for explainability in decision-making~\cite{bibal2020legal}, designers could develop explainable systems to promote self-regulatory efforts while obscuring the need for strict regulation. The current efforts on AI ethics have been largely ineffective and vulnerable to industry manipulation~\cite{resseguier2020ai}. Using XAI as a form of soft regulation may fail to encourage the responsible deployment of decision-making algorithms. We come back to the issue of self-regulation in Section~\ref{sec:hardregulation}, in which we defend hard regulation for AI systems.

It is worth noting that we do not argue that designers should \emph{always} be held solely responsible for the harms caused by decision-making algorithms. Other actors may also be responsible depending on the circumstances. Instead, we demonstrated how designers' power over XAI systems enables them to create the impression that algorithms and patients are to blame for harmful outcomes, impacting how policy decisions are made in the real world. As the famous saying goes: ''with great power there must also come---great responsibility.'' The threshold for shifting responsibility away from designers may need to be higher to ensure they do not escape deserved responsibility. The conflict presented in this paper can be exemplified by the burden of proof that lies with the victims of disparate treatment under US labor law. Designing AI systems that apparently empower patients to prove discrimination through post-hoc explanations shifts responsibility to individuals who do not control or understand algorithmic decision-making.

Although we have focused our discussion on the responsibility of patients and designers, another possible responsible actor is the user of decision-making algorithms. For instance, users could compensate those harmed through the profits they derive from employing AI systems. Interestingly, users remain in a position of both power over patients and subjugation to designers. While those who use decision-making algorithms can employ some of the tactics above to shift responsibility towards patients, they may also absorb responsibility from designers due to their proximity to possibly harmful algorithmic decisions~\cite{elish2019moral}. Future work could explore the responsibility of users vis-à-vis patients and designers. Nevertheless, the difficulty of delimiting the responsibility of distinct actors when machines cause harm calls for proactive approaches that remove any ambiguity, as we discuss below.

%% file: content/4implications.tex
The concerns raised above are mostly backward-looking, focusing on who bears responsibility for the negative consequences of algorithmic decision-making. Responsibility, on the other hand, is not necessarily backward-looking. It can also be forward-looking, emphasizing that individuals should act proactively and responsibly to the best of their abilities to ensure that future outcomes are positive~\cite{van2015moral}. Failure to attend to one's forward-looking responsibilities could lead to the attribution of backward-looking responsibility~\cite{van2011relation}. For instance, if an agent takes the responsibility (in a forward-looking manner) for ensuring that a decision-making algorithm does not discriminate against women, future audits that find this system to favor men make holding the agent responsible in a backward-looking manner appropriate (e.g., by blaming or punishing them).

Researchers have argued that concerns about responsibility gaps can be reduced or even eliminated if designers take responsibility for all stages of the development and deployment of decision-making algorithms~\cite{champagne2015bridging,van2021responsible,johnson2015technology}. However, they overlook the possibility that explainable systems could become a tool that exacerbates these difficulties, particularly (but not exclusively) if designers use them to avoid backward-looking attributions of responsibility.

Building upon previous literature proposing accountability frameworks, we discuss how these approaches could be used to mitigate the potential negative consequences of XAI systems. We build on previous suggestions by recommending concrete steps to ensure that responsibility is not wrongly shifted to patients and algorithms.

\subsection{Accountability for Algorithmic Decision-Making}

While accountability is frequently cited as a backward-looking notion of responsibility~\cite{van2015moral} and as a reaction to blameworthiness judgments~\cite{shoemaker2011attributability}, we now turn our attention to the literature defining accountability as a forward-looking notion with potential backward-looking consequences. One such example is Boven's work~\cite{bovens2007analysing}, which defines accountability as the relationship between an actor and a forum under which the actor is obligated to explain and justify their behavior. The forum poses questions and judges the actor by imposing (backward-looking) consequences, such as punishment. This definition emphasizes the actor's obligation to explain its conduct upon request, implying that actors have a forward-looking responsibility to respond to the forum. Nevertheless, if the forum deems it necessary, this requirement may lead to the imposition of backward-looking responsibilities.

Several principles have been proposed to ensure this accountability relationship can be implemented. One of these principles is traceability, which~\citet{kroll2021outlining} advocates as a critical component in ensuring accountability for algorithmic systems. Traceability is defined as the requirements that should be satisfied so that an algorithm's outputs can be ``understood through the process by which [it] was designed and developed.'' Kroll proposes a set of conditions (e.g., transparency in the design process, reproducibility) and tools (e.g., structured logs) to enable accountability. This approach, however, does not propose a structured framework that designers can readily and explicitly use in the development process.

We build upon Cobbe et al.'s recent proposal of reviewability as a framework for ensuring accountability in algorithmic decision-making~\cite{cobbe2021reviewable}. Because algorithmic decision-making incorporates several human and temporal components, the authors argue that it should not be viewed solely as a technology. Instead, their proposal aims to understand algorithms making decisions as a broad socio-technical process. Human designers, users, patients, and other stakeholders are involved in this process not just during development and deployment but also during conception, investigation, and all other intermediate steps. The reviewability framework maintains accountability by recording ``contextually appropriate information'' throughout the entire socio-technical process of algorithmic decision-making. Below, we explain each step of the framework in-depth and discuss how it could handle the conflict between accountability and explainability put forth in this paper.

\subsubsection{Commissioning}

The first step of the reviewability framework---called commissioning---addresses anything relevant prior to developing the decision-making algorithm. \citet{cobbe2021reviewable} argues that this step should define the problem algorithmic decision-making aims to solve and how it will impact society and individuals. In other words, commissioning refers to all initial human decisions that influence how the model will be developed and used. We argue that one additional and crucial question should be asked: is explainability necessary for the decision-making algorithm under commissioning?

Robbins~\cite{robbins2019misdirected} questioned whether all algorithms, particularly those deployed in low-stakes scenarios, should be explainable and argued against it. We suggest that this inquiry should also examine whether the problem under consideration requires explainability in light of the responsibility issues discussed above.

We also propose that interpretable systems should be prioritized over those that give post-hoc explanations. As discussed above, post-hoc explanations can lead to further problems concerning perceived responsibility and control. In contrast, interpretable systems place a greater emphasis on the roles of designers in developing decision-making algorithms because these systems become explainable not only after deployment but also during the development process. If post-hoc explanations are necessary for a specific problem, designers should consider which types of explanations should be provided. Explanations should not highlight the algorithm's agentic role in the process, dealing with people's perception that AI systems should be held responsible. Furthermore, explainable algorithms should genuinely empower patients rather than instilling a false sense of confidence and control.

\subsubsection{Model Building}

The second step of the reviewability framework focuses on the technical components of algorithmic decision-making. \citet{cobbe2021reviewable} claims that maintaining accountability requires knowledge about how data is collected and pre-processed, how the model is trained and tested, and how and by whom relevant decisions are made. Considerations about XAI and responsibility are also important throughout the model building process. Explainable systems may require different datasets and pre-processing. Although some types of explanations do not require specific datasets or annotations (e.g., feature importance explanations can be generated without additional data), others (e.g., natural language explanations) may require specific data for training. Datasets should be collected and pre-processed in ways that do not conflict with designers' forward-looking responsibilities.

Model testing is one of the most important steps in the reviewability framework to avoid any conflict between XAI and responsibility. Explainable systems should be tested rigorously with those who will use and be subjected to them once deployed. The benefits are numerous: understanding the stakeholders' view (i.e., how they perceive explainable algorithms) assists designers to prevent AI systems from being regarded as rational and intentional, which could lead to the belief that they should be held responsible; studying how users and patients interact with these systems can prevent overreliance (as done in~\cite{buccinca2021trust}) and a false sense of control over algorithmic decision-making; testing and improving explanations can ensure they are compatible with legal and social requirements. To ensure that responsibility for algorithmic decision-making is distributed fairly, users and patients must be included in the validation of explainable systems.

\subsubsection{Deployment \& Investigation}

The final two steps of the reviewability framework encompass the operation of decision-making algorithms (e.g., how systems are deployed, supported, and their consequences) and any subsequent investigative activity (e.g., internal and external audits). Here, users and patients play a significant role since they are intimately involved in any potential outcomes. As previously stated, these actors may also be held responsible for algorithmic decisions. Designers, for instance, cannot stop users from using algorithms for nefarious purposes and should not be held liable if users use them for unforeseeable wrongful actions. Patients may exploit explanations and abuse the system, which may be immoral or illegal.

Nevertheless, we highlight that explainable systems require particular attention concerning how responsibility should be distributed in both forward- and backward-looking manners. The deployment of XAI should not delegate responsibilities that should have been considered in previous steps of the framework to users and patients. These actors do not have complete control over decision-making algorithms and their outputs, even if explainability aims to empower them. Responsibility for algorithmic decision-making should be assigned proactively during the initial steps of the accountability framework. When the framework's final steps come into play, the respective forward- and backward-looking responsibilities of designers, users, and patients should be explicit and obvious. Such a clear division of responsibility should not be self-regulated without legally binding forces; instead, it should be codified into law.

\subsection{The Necessity for Hard Regulation}
\label{sec:hardregulation}

We have advocated for assigning forward-looking responsibilities to designers throughout this work to avoid the conflict between explainable and accountable algorithmic decision-making. By emphasizing the prospective aspects of responsibility, designers can be held accountable for the harmful outcomes of decision-making algorithms, mitigating incorrect backward-looking attributions of responsibility to patients and algorithms. This approach underlines the non-technical components of algorithmic design, particularly during the commissioning and testing steps of the accountability framework presented above.

We have extensively discussed the literature questioning whether designers would take responsibility for their systems and argued that explainability might be used to escape responsibility. This would be comparable to what Floridi called ``ethics washing''~\cite{floridi2019translating}, in which actors engage in allegedly ethical behavior to evade regulation. What is needed to ensure that designers are held accountable for algorithmic decisions? We contend that it is hard regulation.

Jobin et al.~\cite{jobin2019global} found explainability to be the most mentioned principle in a review of AI ethics guidelines and principles put forward by industry actors, policymakers, and academics. There appears to be an overreliance on ethics to ensure that responsibility is not corrupted; however, principles alone cannot guarantee ethical AI~\cite{mittelstadt2019principles}. Ethics is not a substitute for hard regulation as it can be easily exploited by powerful actors and cannot ensure that principles are followed~\cite{resseguier2020ai}. Self-regulation and soft law appear to face similar problems.

Government regulation proposals currently do not address the conflict between accountability and explainability. The European Union (EU) AI Act, the most recent attempt to regulate AI systems, does not directly address any of the responsibility issues raised by autonomous machines. Although EU expert groups have proposed potential liability models for AI systems~\cite{euliability} and the EU Commission has debated revising product liability laws~\cite{euproduct}, the AI Act appears to allow designers to avoid liability rather than addressing how and when they will be held liable. The current proposal allows designers to ``wash their hands'' by adhering to local standards through self-assessment protocols. Most crucially, these local standards are susceptible to lobbying by private organizations. By complying with private-sector standards and meeting regulatory transparency requirements, designers could shift perceived responsibility for negative outcomes to patients, who lack the necessary understanding and control over algorithms. This approach disregards the overwhelming power that designers have over XAI systems in comparison to other actors.

If responsibility frameworks are codified into law, designers will not be able to shift responsibility to other stakeholders or algorithms. Elucidating which roles and obligations designers, users, and patients should have could help mitigate the problem of the former falsely empowering others to escape blame, punishment, and other forms of backward-looking responsibility. Such a legal framework could even hold actors jointly liable if they fail to meet their obligations~\cite{vladeck2014machines}. While it is beyond the scope of this paper to argue for a specific legal framework, we highlight the need for more research on the subject in the future. Explainable decision-making algorithms, for instance, may be required to declare their lack of agency, intentionality, and rationality alongside their explanations so that people are not influenced to hold them accountable. Such an approach would be similar to the existing proposals that mandate designers to disclose bots~\cite{veale2021demystifying}. In conclusion, regulation can ensure that explainability and accountability coexist in algorithmic decision-making.

\subsection{Blameworthy Algorithms?}

People ascribe blame, responsibility, and punishment to algorithms upon harm (e.g.~\cite{lima2021human,lima2021punish,furlough2019attributing,malle2015sacrifice,malle2019ai}). In this paper, we argued that explainable systems may be attributed even higher levels of backward-looking responsibilities when they provide post-hoc explanations. This effect raises the question of whether an algorithm can be held responsible for its actions. Although most academics agree that responsibility (such as blame and punishment) may be inappropriate for algorithmic systems (e.g., see~\cite{danaher2016robots,sparrow2007killer,bryson2017and}), others approach this question through a different lens~\cite{gunkel2017mind,stahl2006responsible,coeckelbergh2009virtual}. Given the empirical evidence that humans may attribute different notions of responsibility to these systems~\cite{lima2021human}---even though they are aware that doing so is unfeasible~\cite{lima2021punish}---we suspect that holding algorithms responsible may take two different routes.

One path dismisses any prospect of holding these systems responsible, arguing that the responsibility for algorithmic decision-making should be left to humans and not algorithms~\cite{van2021responsible}; doing otherwise could lead to vast social and legal unrest~\cite{solaiman2017legal}. The narrative that AI systems should be held responsible may dilute the much-needed responsibility of designers~\cite{bryson2017and}. To progress along this path, AI systems should be built to refute incorrect attributions of agency, intentionality, or responsibility to algorithms. As mentioned above, explainable systems could highlight their lack of agency when providing explanations. Algorithms could be designed to oppose any attribution of mind, which has been shown to influence people's behavior towards them~\cite{lee2019s,de2014importance}.

A more controversial approach would be to invest in holding algorithms responsible, although not to the same extent or in the same way that humans are held accountable. \citet{coeckelbergh2009virtual} defended holding such systems virtually morally responsible to the extent that they appear to be moral agents. Similarly, \citet{stahl2006responsible} proposed the concept of quasi-responsibility, under which algorithms could be held responsible for fulfilling social goals; 
according to the author, doing so could facilitate future attributions of responsibility to human stakeholders. AI systems could also be held legally responsible if they are granted legal personhood~\cite{turner2018robot}. These proposals raise the question of whether such a path is feasible and helpful. Is there any benefit in holding algorithmic systems accountable in general, or is it solely detrimental to moral, legal, and social consistency?

%

%% file: content/5conclusion.tex
This paper shared a concerning viewpoint that the current call for over-emphasizing explainability in algorithmic decision-making may conflict with accountability. Based on philosophical interpretations of moral responsibility and empirical research, we suggested that explainable algorithms could be seen as blameworthy and responsible. We also showed how XAI systems could shift perceived control over algorithms away from designers and towards patients and argued that this shift is mistaken. Expanding on past work questioning designers' willingness to take responsibility, we showed how algorithms and patients could become moral scapegoats that might absorb the responsibility of designers.

To avoid the potential conflict between explainable and accountable algorithmic decision-making, we have argued for a greater emphasis on designers' forward-looking responsibilities. Existing accountability frameworks should be modified to include explainability considerations at every step of AI development and deployment. XAI is an important part of the responsible deployment of algorithmic decision-making, but it should not be viewed as a panacea to all problems. The XAI field is critical for improving algorithmic decision-making, and society should be aware of how those in power may abuse it.